\documentclass[journal,onecolumn,12pt,twoside]{IEEEtranTCOM}
\usepackage{amsmath}
\usepackage{xcolor}
\usepackage{hyperref}
\normalsize
\usepackage[cmintegrals]{newtxmath}
\usepackage{cite}
\ifCLASSINFOpdf
  \usepackage[pdftex]{graphicx}
\else
\fi
\usepackage{url}
\usepackage{geometry}
\begin{document}
\title{Towards the Use of Re-configurable Intelligent Surfaces in VLC Systems: Beam Steering}
\author{Alain R. Ndjiongue, Telex M. N. Ngatched, Octavia A. Dobre, and Harald Haas
\thanks{\textit{Corresponding author: AR Ndjiongue, ndjiongue@gmail.com}}} 
\markboth{}
{}
\maketitle
\begin{abstract}
The emergence of visible light communication (VLC) technology as a solution to solve radio frequency impediments, such as spectrum shortage, is continuously appealing. In addition to its large and unlicensed bandwidth, VLC provides a high level of security in a closed room with zero radio frequency interference. However, loss of the VLC signal is experienced when the receiver rotates or moves. This challenge requires special solution for the integration into portable devices. On the other hand, re-configurable intelligent surface (RIS) is a technology exploited in radio frequency to solve dead zones and loss of signal. RIS elements are characterized by tunable physico-chemical characteristics including physical depth and refractive index. In this paper, we exploit these RIS attributes to steer the incident light beam and offer the VLC receiver a large range of rotation angle, while improving its signal detection capabilities. We show that instead of using convex, parabolic, or spherical lenses, adopting a meta-lens with artificial muscles or a thin-film crystal-liquid with embedded Titanium dioxide nano-disk, a VLC receiver can detect light rays at $90^o$ incidence angle with a high precision and considerable improvement in the detected light intensity, even with a miniaturized single photodetector. 
\end{abstract}
\IEEEpeerreviewmaketitle
\section{Introduction} \label{intro}
Visible light communication (VLC) technology has attracted tremendous attention in recent years. It is progressively getting matured and its mass production will happen in a near future. It is the preeminent communication technology used in light fidelity, and presents outstanding advantages in indoor positioning and intelligent transportation systems \cite{7901496, 8970387}, to mention only a few. Despite all its advantages over radio frequency, which comprise higher and unregulated bandwidth, and cost-effectiveness, the VLC technology faces some drawbacks including short transmission range and loss of signal which may occur when the receiver rotates or moves. When the VLC signal is generated by light emitting diodes, there is a trade-off between coverage and distance which can be controlled by field-of-view (FoV) of the LED. Optical receiver systems also have a FoV and the received data-carrying light photons must fall within the receiver FoV. The latter, in conventional VLC receivers, is determined by the size of both lens and photodetector (PD) \cite{7498569}. Most VLC receivers use convex, spherical, or compound parabolic concentrators \cite{8417543, 8937517, Wang}, to focus the incoming light beam to the PD's surface. The more data-carrying photons arrive at the PD, the higher the received energy, and the better the bit-error-ratio performance and data-rate. This suggests that a large PD is advantageous. However, the larger the PD, the smaller its bandwidth, which is not desirable as the data-rate scales linearly with bandwidth. Therefore, the optimum system employs a very small detector, and uses a system that redirects as many photons as possible into the small-area PD. This paper tackles this challenges using re-configurable intelligent surfaces (RISs).   

The RIS technology is new and is attracting significant interest in research \cite{8796365,9133588}. This is due to its capability to build a programmable wireless environment for future communication systems. The technology was first introduced by Berry in 1963, where he proposed the reflect-array antennas \cite{4653662}. Since then, the research community has studied absorbing boards, selective windows and walls, and frequency selective surfaces. Nowadays, RIS bears several other names. The concept is called: (\textit{i}) large intelligent surface to indicate the area exploited to contain the RIS units; (\textit{ii}) large intelligent meta-surface or re-configurable meta-surface because the RIS element is characterized by a complex and artificial electromagnetic structure; (\textit{iii}) intelligent reflecting surface or smart reflect-arrays because the incident signal may be reflected on the RIS surface; (\textit{iv}) passive intelligent surface or passive intelligent mirrors to specify that the RIS elements do not amplify the incident signal. The main advantage of RIS infrastructures is the electronically maneuverability of their physico-chemical characteristics, to benefit phenomena such as reflection and refraction. Present research trends mostly focus on exploiting the RIS reflective characteristics to solve death zones and improve signal quality \cite{8796365}. In contrast, in this paper, we analyze its refractive attributes to steer the light beam and improve signal reception in VLC systems.
\begin{figure*}
	\centering
	\includegraphics[width=0.95\textwidth]{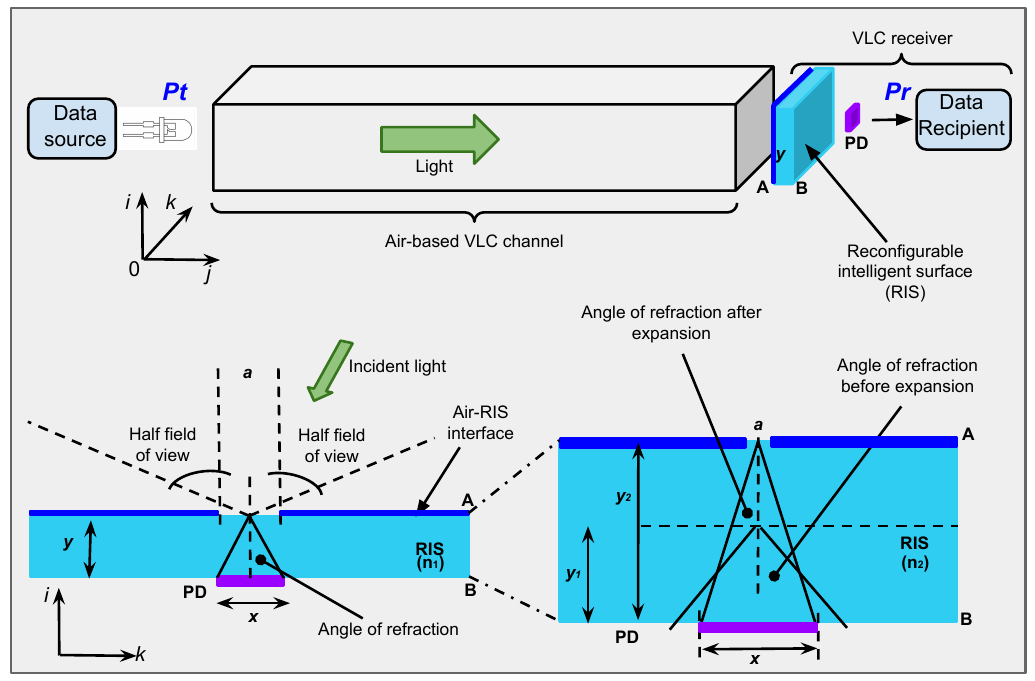}
	\caption{Descriptive diagram of a RIS-based VLC communication system. Top: Diagrammatic representation of the full VLC system using RIS; bottom-left: Cross-section of the RIS-based receiver; bottom-right: Stretched RIS in a VLC receiver, showing a different refractive index and RIS physical depth.}
	\label{fig:model}
\end{figure*}

In VLC systems, signal loss and poor detection are primarily observed in long-distance and non-line-of-sight transmissions, but also in short-range line-of-sight communications when the PD is not oriented towards the incident light. In the latter, signal loss occurs when the transmitted light rays do not fall in the receiver's FoV \cite{7809099}. Commonly, VLC receivers are equipped with a convex lens, which helps to focus the incident light rays to the PD's surface. Hence, its main role is to steer the incoming light beam in order to direct light rays, which are falling in the receiver's FoV, to the PD's surface. This helps to lessen the signal loss and improve the detection performance. In general, steering the VLC light beam is performed at the receiver level, between the lens and PD. Except in case where many lenses are exploited, for example in camera receivers, these refractive traditional convex lenses are static with fix characteristics. An image receiver using a convex lens provides about 2 mm spot size with a FoV up to 85$^o $ \cite{8417543}, while a gradient index lens with 9 mm width and 15 mm height, and compound parabolic concentrator yields a FoV of about 40$^o $ \cite{Wang}. A catadioptric-monolithic bi-flat and bi-parabolic lens of size 46 mm can concentrate the incident light from a FoV of 85$^o $ to a miniaturized spot; however, it presents a loss of intensity when the incidence is greater than 25$^o $. A design of spherical optical receiving antennas for VLC systems based on the Taguchi method provides a minimum spot of 3 mm with a 45$ ^o $ FoV and a spherical diameter of about 6 mm \cite{8937517}. The above state-of-art of VLC lens does not have the capability to dynamically steer the incident light beam, and hence limits the receiver's detection capabilities, especially when it rotates or moves. 
\begin{figure*}
	\centering
	\includegraphics[width=0.95\textwidth]{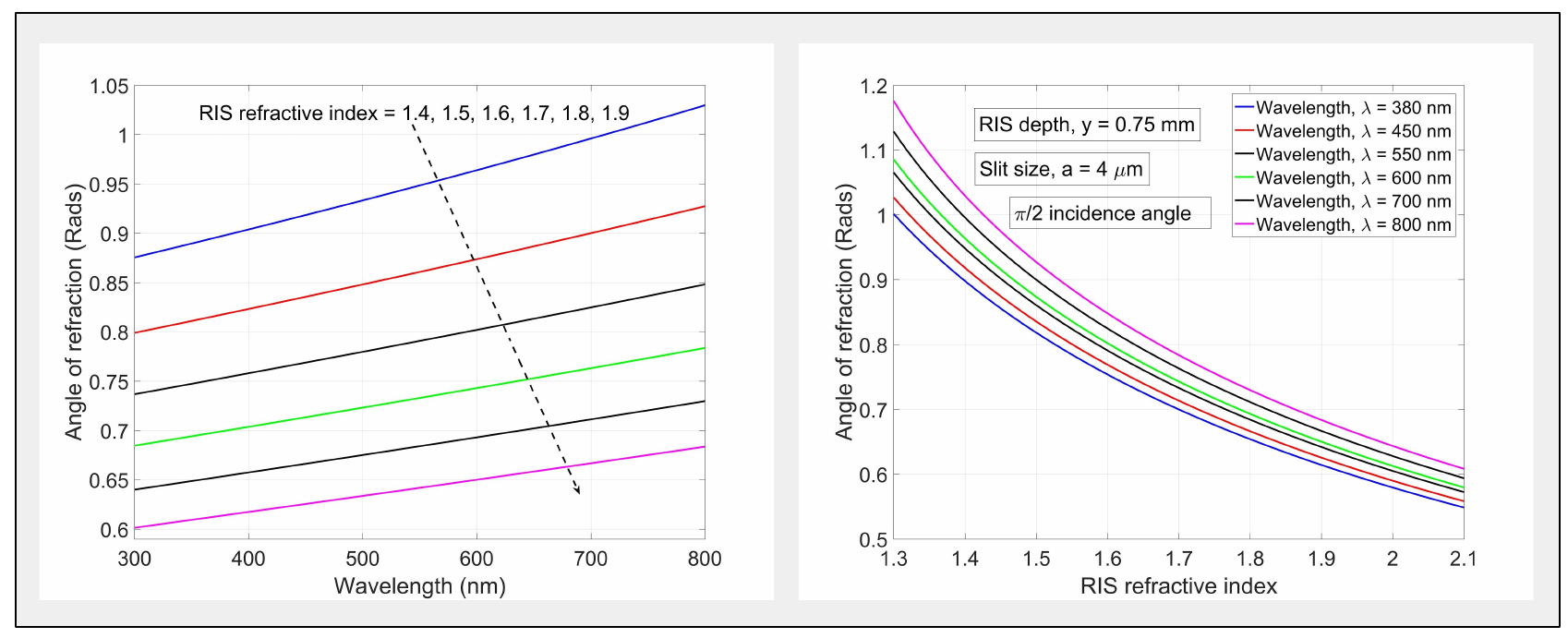}
	\caption{Steering effects on the incident light due to varying the RIS refractive index (left) and stretching a meta-lens (right), for a slit size of 4 $ \mu $m, a RIS physical depth, $ y $ = 0.75 mm, first-order grating ($ m $ = 1), and $90^o $ incidence angle.}
	\label{fig:expansion}
\end{figure*}

In modern optical receivers, several methods may be used to steer the incoming beam. These include and are not limited to: (\textit{i}) stretching a flexible matrix containing dielectric nano-resonators; (\textit{ii}) changing the phase of an amorphous crystalline transition in a chalcogenide; (\textit{iii}) ultra-fast switching of mie-resonant silicon nano-structure; and ($ iv $) adjusting the structure refractive index. However, all these methods have a weak effect on the refracted beam, even at high optical intensity \cite{komar}. Introducing a RIS element, with electronically tunable physico-chemical characteristics, which can be obtained by varying the temperature of a nano-cell or re-orientating crystal-liquid alignment based on induced electrical field, will have a considerable impact on the received beam and signal strength. Equipped with a RIS element, any signal arriving with an incidence angle between 0 and $90^o $ can be steered in such a way that its refracted version falls on the PD's surface, with a miniaturized spot of about 0.1 mm. 

This paper focuses on the use of two types of intelligent meta-elements to steer the incident light beam in VLC systems; a meta-lens with electrically stretchable artificial muscles and a crystal-liquid based RIS infrastructure with electronically adjustable refractive index. They can efficiently replace the traditional lens. A meta-lens with electrically stretchable artificial muscles performs refracted beam manipulation through expansion and compression of the RIS element, while the crystal-liquid based RIS exploits crystal-liquid alignment and orientation to dynamically tune the RIS refractive index. To the best of our knowledge, using RIS elements in VLC technology to steer the incident light beam has not yet been proposed in the open literature, and represents the motivation of this paper. It points out that instead of the ordinary lens, using a meta-lens with stretchable artificial muscles or a crystal-liquid based RIS structure to steer the incoming light rays over the PD's surface, presents tremendous advantages including a large incident angle, which allows the receiver to rotate within a specific range while still being able to detect the incoming signal. The main goal of this paper is to show that using RIS in a VLC receiver has the potential to significantly improve its signal detection capabilities. Toward this end, we make the following contributions: ($ i $) we propose the principle of RIS infrastructure based on both a stretchable meta-lens with artificial muscles and thin-film crystal-liquid with embedded Titanium dioxide (TiO$ _2 $) nano-dicks. We analyze these and provide results showing their impact on the incident light rays; ($ ii $) we examine these items separately, consider their physico-chemical characteristics, and suggest a comparison between them and the traditional convex lenses, to highlight their advantages; ($ iii $) finally, we highlight the advantages related to the use of RIS in VLC systems, and suggest future research directions. 
\section{RIS-based VLC Beam Steering}
\subsection{VLC Channel Steering: The Concept}
The VLC channel is shaped by the transmission environment. In outdoor, objects such as cars, buildings, and trees, placed on the light rays reshape the signal direction depending on their physical characteristics and architecture, while in indoor the contribution of wall, ceilings, floors, and other objects in the room, are undeniable. As a result, the received signal is a sum of line-of-sight and non-line-of-sight rays. Based on the receiver FoV, a specific amount of light reaches the PD through a lens, whose main role is to steer the incident light. Normally, the lens is part of the receiver and is not considered as part of the channel. However, all components acting on the transmitted signal before it reaches the PD may be considered as part of the channel. The convex lens makes use of its shape and physico-chemical properties to guide all incident light rays so that they reach the PD's surface. This concept is also been used in the photography industry and is exploited in equipment such as microscopes. This approach is depicted in Fig.~\ref{fig:model}, where the traditional lens is replaced by a RIS element.
\begin{figure*}
	\centering
	\includegraphics[width=0.95\textwidth]{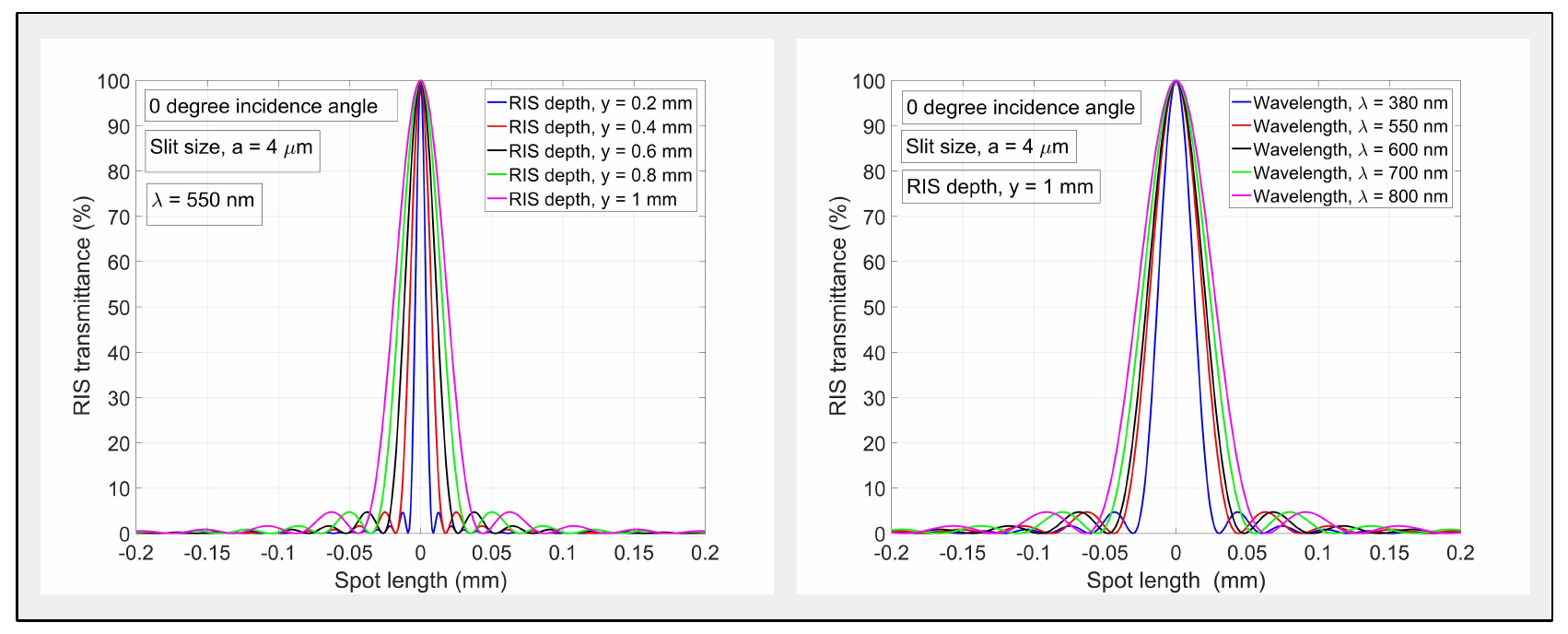}
	\caption{Steering effects on the refracted light concentration due to stretching a meta-lens (left) and varying the wavelength (right), for a slit size of 4 $ \mu $m. Left: 0$ ^o $ incidence angle and RIS physical depth, $ y $ = 1 mm; right: 0$ ^o $ incidence angle and wavelength, $\lambda$ = 550 nm.}
	\label{fig:results2}
\end{figure*}
\subsection{RIS-based VLC Beam Steering}
The top-part of Fig.~\ref{fig:model} depicts a VLC transmission system, in which the lens located inside the receiver, is replaced by a meta-element. It also illustrates the channel, which is the space between the light sources (light-emitting diodes) and the PD, also placed inside the receiver. The figure highlights the fact that the channel is made of the air-based part and the section from the RIS element to the PD. Controlling the VLC light beam can be performed by tuning the shape of light through adjusting the geometry of the transmission environment, or a part of it. Given that it is not easy to control the shape of the room, this control can efficiently be done inside the receiver using meta-elements. This can be achieved by varying the RIS thickness, $ y $, or its refractive index, $ n_{RIS} $. Hence, the physical dimensions of the RIS element ($ y $) or its chemical structure ($ n_{RIS} $) can be tuned to control and steer the incoming VLC light beam. This is shown in the bottom-part of Fig.~\ref{fig:model}, which displays a cross-sectional representation of the RIS element. Its upper surface is covered by a thin-film of perfect optical opacity material with a centralized slit of dimension $ a $, and the detector is a pastille of length $ x $ centrally placed at the bottom of the RIS structure. Note that the cross-sectional shape of the RIS structure is not important. A monochromatic and plane light-wave of wavelength, $ \lambda $, falling on the RIS surface with an incidence angle, $ \theta_a $ (half of the receiver's FoV), is diffracted into different orders, $ m $, inside the RIS structure and under a refraction angle, $ \theta_{RIS} $. This forms a refraction system, which is governed by the grating equation. This equation relates the above parameters to the receiver's FoV and the air refractive index, $ n_a $ \cite[Eq. (37.1)]{9131443}, and is used to predict the different directions of light propagation from a grating in a diffraction phenomenon.

The VLC beam steering, discussed in the rest of this paper, based on both refraction and diffraction, uses the first-order grating, $ m = 1 $, as it provides the strongest signal intensity \cite{9131443}, and satisfies the principle of etendue. The refraction phenomenon stipulates that when light travels from one substance to another, its speed is adjusted and as a result, its direction changes. On the other hand, diffraction, which occurs when a wave encounters an obstacle provided with a slit, affects the light intensity after its course in the second substance.  

\textbf{Influence of the RIS refractive index: } The incident light, falling on a single slit, is diffracted and the refracted light appears as from a new source represented by the single slit. Direction and intensity of the refracted light mainly depend on the RIS refractive index, slit size, RIS physical depth, and incidence angle. Figure~\ref{fig:expansion}, which plots the first-order grating equation \cite[Eq. (37.1)]{9131443}, depicts effects of the RIS refractive index on the refracted light. Analysis is performed for the maximum incidence angle ($\theta_a = 90^o $) and a RIS physical depth, $ y $ = 0.75 mm. The left-hand-side of Fig.~\ref{fig:expansion} shows that the angle of refraction is proportional to the wavelength and varies considerably with the refractive index. As an example, for a wavelength, $\lambda$ = 300 nm, at a refractive index of 1.4, the angle of refraction is 50.133$ ^o $. This value drops to 34.377$^o $ when the refractive index changes from 1.4 to 1.9. Note that at a wavelength of 300 nm, we obtain a variation of the refraction angle close to 14.725$ ^o $ for a refractive index variation of 0.5, while for the same refractive index discrepancy, we obtain a change of 20.053$ ^o $ in the angle of refraction for a wavelength close to 800 nm. These results are confirmed by the right-hand-side of Fig.~\ref{fig:expansion}, where the angle of refraction is given in terms of the refractive index, for multiple values of the wavelength. They suggest that tuning is more noticeable for higher values of the wavelength. 
\begin{figure}
	\centering
	\includegraphics[width=0.45\textwidth]{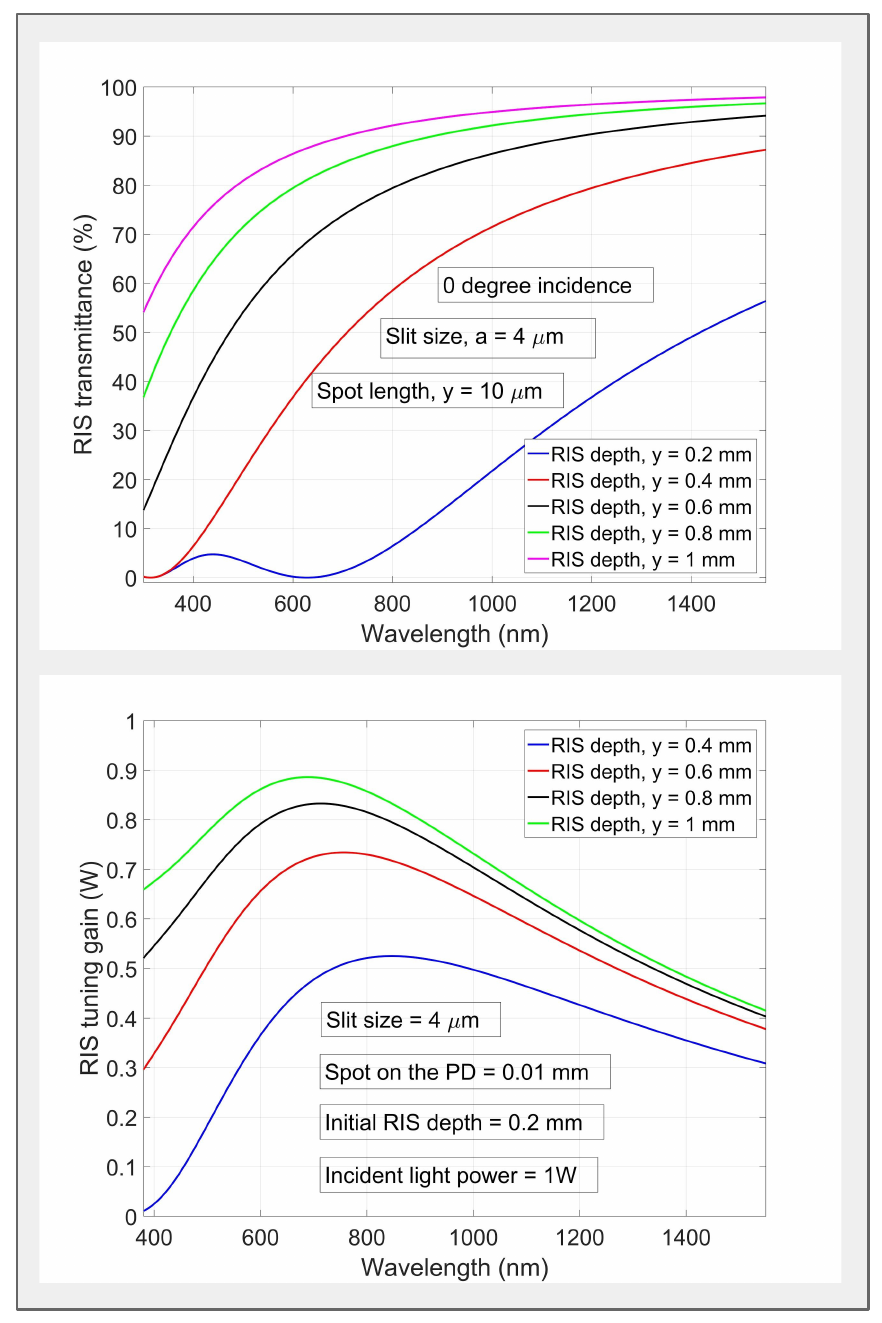}
	\caption{Tuning gain in terms of wavelength for different RIS physical depths and a slit size of 4 $ \mu $m, a 0$ ^o $ incidence angle and an incident light power of 1 W. Top: relative transmittance of the RIS structure for different values of $ y $; bottom: stretching gain for different stretching levels considering an initial physical depth of $ y $ = 0.2 mm.}
	\label{fig:gain}
\end{figure}
\begin{figure*}
	\centering
	\includegraphics[width=0.94\textwidth]{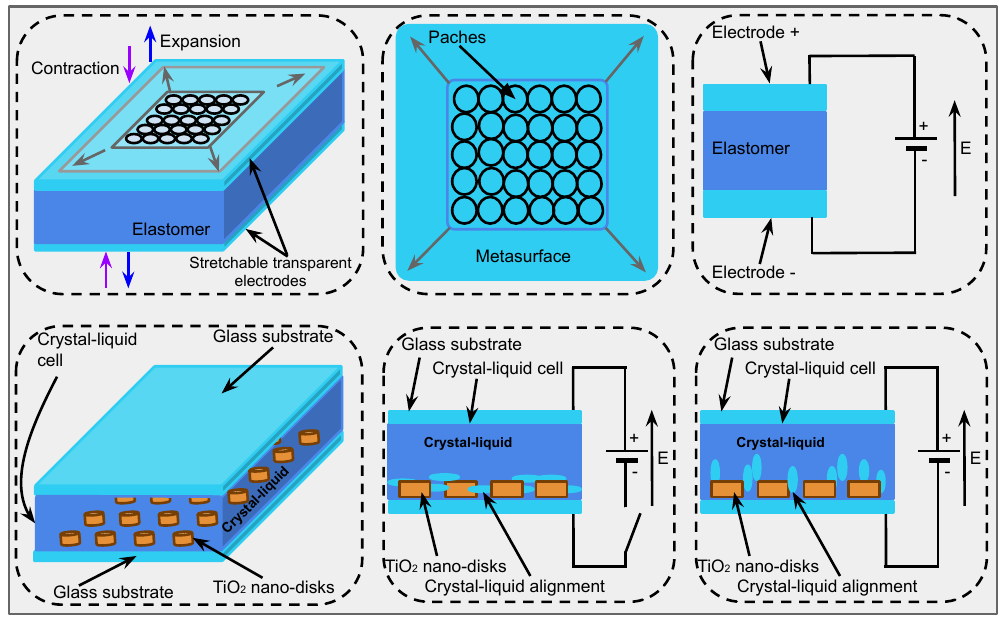}
	\caption{Sketch of (top): A meta-lens with an artificial muscle, its perspective and top views, and the stretching process by electrical tuning; (bottom): A crystal-liquid-based RIS with embedded TiO$ _2 $ nano-disks and its electrically tuning principle.}
	\label{fig:RISVLC}
\end{figure*}

\textbf{Effects of the RIS physical depth: }The importance of using a single slit to create a diffraction in a VLC receiver is to readily control the light spot and intensity on the PD's surface. This summarizes the effects of RIS physical depth, $ y $, on the refracted light, which is twofold. It simultaneously acts on the light intensity and spot size on the PD's surface. On the left-hand-side of Fig.~\ref{fig:results2},\footnote{Results in this figure are obtained based on the the single slit Fraunhofer diffraction pattern \cite[Eq. (37.2)]{9131443}.} the RIS transmittance is given in terms of spot length for multiple values of $ y $ and a light falling perpendicularly on a slit of length 4 $ \mu $m. This figure shows that the spot size is a function of the RIS physical depth for a given wavelength, slit size, and angle of incidence.

\textbf{Influence of wavelength: } Different impacts of $\lambda$ on the refracted light are already demonstrated in Fig.~\ref{fig:expansion} and on the left-hand-side of Fig.~\ref{fig:results2}. The right-hand-side of Fig.~\ref{fig:results2}\footnotemark[1] shows that the spot size also varies with the wavelength for fixed values of the incidence angle, slit size, RIS physical depth, and refractive index. 
\subsection{Tuning Gain}
\textbf{Impact of the incidence angle: } The transmittance of an RIS element, defined as the ratio of light intensity at the slit by that of the PD's received signal, is mostly impacted by the angle of incidence. This is due to the fact that the intensity applied to the slit is multiplied by $\cos(\theta_a)$ to obtain a wave perpendicular to the slit, as shown in Fig.~\ref{fig:results2}, obtained for a 0$ ^o $ incidence angle. The value of transmittance at the PD's center would be less than 100\% if the incidence angle is greater than 0$^o $. This figure also shows that for any value of $ m $, the refracted light intensity increases with the decrease of the incidence angle.

\textbf{Effects of the RIS physical depth: } The right-hand-side of Fig.~\ref{fig:results2} shows that the refracted light intensity varies with the wavelength. However, the impact of the RIS physical depth is considerable, as confirmed by the top-part of Fig.~\ref{fig:gain}.\footnotemark[1] Around a wavelength of 600 nm, a RIS structure with physical depth of 0.2 mm totally absorbs the incoming light, while its variation of about 0.8 mm reduces this absorption by about 90\%, to provide almost the totality of incoming light intensity to the PD. This part of Fig.~\ref{fig:gain} also shows that at high values of wavelength, varying the RIS physical depth has no impact on the incoming light intensity as all curves converge to 100\%. 

The tuning gain, which is due to the conservation of basic radiance, is defined by the difference between transmittance before and after expansion. It tells how much intensity is gained or lost after RIS expansion or compression. It also reveals the wavelength at which a better transmittance is obtained for a specific type of RIS element. This parameter is presented in the bottom-part of Fig.~\ref{fig:gain}\footnotemark[1] for a signal falling on a 10 $ \mu $m of spot length, an initial RIS physical depth of 0.2 mm, a 0$ ^o $ incidence angle, and a slit size of 4 $ \mu $m. This part of Fig.~\ref{fig:gain} also reveals that there is an intensity gain obtained from the tuning operation, which vary with the RIS physical depth and wavelength. Further, note that changing the slit size changes the wavelength at which the peak transmittance occurs. This parameter is important in designing meta-lenses with artificial muscles given that its physical size changes during tuning.   
\section{RIS Design for VLC Systems}
We consider two different types of material and structure in designing RIS elements for VLC systems; a meta-lens with stretchable muscles and a crystal-liquid based RIS element. The design complies with the principle of beam parameter product to ensure that the etendue of the light refracted and basic radiance are conserved. Figure~\ref{fig:RISVLC} displays a sketch of the two meta-element types and their electrically tuning principle. In both cases, the design of the element is motivated by two main transmission aspects: steering the light beam and improving the received light intensity. Based on a first-order grating, using a single slit diffraction, and considering the highest incidence angle, $90^o $, the refractive index of the RIS element is determined such that all rays falling on the slit are directed to the PD's surface.
\begin{table*}
	\begin{center}
		\caption{Comparison between traditional lenses and dynamically tunable meta-elements.}
		\label{tab:table1}
		\colorbox{lightgray!25}{
			\begin{tabular}{|l|c|c|c|c|c|c|c|}
				\hline
				& Tuning Flexibility & Dynamic Control & Tuning Gain & Incidence Range & Complexity & Ext. Voltage\\
				\hline
				Convex lens & // & // & // & $\leq 36.2^o $ & Average & No\\ \hline
				GILCPC lens & // & // & // & $\leq 40^o $ & Average & No\\ \hline
				Spherical lens & // & // & // & $\leq 45^o $ & Average & No\\ \hline
				CMBBP lens & // & // & // & $\leq 85^o $ & Average & No\\ \hline
				Adj-lens & Low-avg & Yes & Yes &$< 90^o $& Avg-high & Yes/No\\ \hline
				ML-AM & High & Yes & Yes &$\approx 90^o$& Avg-high & 1 kV - 3 kV\\ \hline
				CL-RIS & High & Yes & Yes &$\approx 90^o $& high & 2 V - 5 V \\ \hline
				\multicolumn{7}{c}{GILCPC = gradient index lens with compound parabolic concentrator; CMBBP = catadioptric monolithic bi-flat}\\ \multicolumn{7}{c}{bi-parabolic lens; Adj-lens = lens with adjustable elements; ML-AM = meta-lens with artificial muscles;}\\
				\multicolumn{7}{c}{CL-RIS = crystal-liquid based RIS; Ext. = external.}\\
		\end{tabular}}
	\end{center}
\end{table*}

\textbf{Refractive index: }The RIS refractive index determines the angle of refraction for a specific incidence angle. Tuning this parameter smoothly provide the structure with a largely modulated light beam through the RIS structure. Some values obtained by analysis are between 1.508 and 1.7539. These refractive indexes correspond to materials that can be used to produce the same result for different wave frequencies, and distinct expansion of the meta-element. They are obtained for an initial RIS thickness of 0.75 mm, a slit size of 100 $ \mu $m, and a targeted spot of 1 mm, for the maximum incidence angle (90$ ^o $). A few materials bearing these indexes are pure flint glass ($ n $ = 1.6 to 1.62), impure flint glass ($ n $ = 1.5 to 1.755), pure crown glass ($ n $ = 1.50 to 1.54), impure crown glass (1.485 to 1.7555), and thin-film crystal-liquid with embedded nano-disks (1.508 to 1.9 depending on the temperature conditions) \cite{refrac}. 

\textbf{RIS physical depth: }The RIS physical depth defines the thickness of RIS elements, informs on how much the incident signal can be refracted, and determines the intensity gain as well. It also specifies the size of light spot on the PD's surface. 
\subsection{Example of Meta-lens Design} 
Figure~\ref{fig:RISVLC} shows a design of a single-cell RIS structure. Its top-part is a meta-lens with stretchable muscles. On the left, it shows the perspective view, the top-view in the middle, and the electrically tuning process on the right. The meta-element is made of a meta-surface layer containing graphene patches. This structure seats on two parallel stretchable electrodes, which contain an elastomer in between, and is characterized by the flexibility of the meta-surface. This surface's perimeter can increase or reduce, while its thickness is easily compressed or expanded. These operations have an effect on the direction, amplitude, phase, and wavelength of incident light rays, which can therefore be selected depending on the required light color. Applying a voltage on the stretchable electrodes (top-right-hand-side of Fig.~\ref{fig:RISVLC}) causes their expansion or compression, which helps to redirect incident lights by modifying the physical dimensions of the material. This affects emission and absorption coefficients, and the refraction index. Practically, high voltages should be applied to stretch the electrodes. This can be controlled by an efficient algorithm for the purpose of accuracy and efficiency. The example shown in the top-part of Fig.~\ref{fig:RISVLC} is a converging square meta-lens made of two thin contractible-expansible transparent electrodes sandwiching a dielectric elastomer actuator. The total structure, which is about 30 $ \mu $m thick, can be expanded to 15 mm, and is capable of converging a laser-light between 1440 and 1590 nm for a spot on the PD of size from 21.4 $ \mu $m to 37.7 $ \mu $m, under an applied voltage between 0 and 1 kV \cite{she2018adaptive}, when the upper surface is made of an optically opaque plate with a centralized slit of 4 $ \mu $m length, and the PD located at 50 mm from the meta-lens.
\subsection{Example of Crystal-Liquid Design}
The bottom-part of Fig.~\ref{fig:RISVLC} depicts a single-cell RIS structure based on a thin-film crystal-liquid and its tuning principles, showing its perspective-view, no-power, and maximum applied voltage scenarios. The structure is made of a TiO$ _2 $ with crystal-liquid infiltration sandwiched by two glass subtracts. It uses a crystal-liquid cell of 2 $ \mu $m thickness, TiO$ _2 $ nano-disks characterized by 320 nm diameter, 200 nm height with a 250 nm spacing, and glass substrates of 30 $ \mu $m thickness. The structure can deviate a light from a maximum of incidence angle ($90^o$). To achieve this, the refractive index needs to be tuned with a difference of about 0.2 to 0.4, when the structure is powered by a source with voltage between 3 V and 5 V, in a strong room temperature \cite{sun2019efficient}. The structure requires an initial preparation of all molecules, to align them in parallel to the incident light polarization direction. Note that this alignment is proportional to the source's voltage. A variation of the input power causes a re-alignment of the crystal-liquid inside the cell, resulting in a modification of the refractive index. While varying in the same direction as the applied voltage, this refractive index defines the incident signal orientation and also has an impact on RIS transmittance \cite{komar, sun2019efficient, she2018adaptive}. 
\subsection{Comparative Analysis}
In this subsection, we offer a comparison of traditional lenses to electronically tunable meta-elements. This is depicted in Table~\ref{tab:table1}, which displays steering parameters of single convex, parabolic and spherical lenses, a convex lens with adjustable elements, meta-lens with artificial muscles, and crystal-liquid based RIS. These steering parameters are tuning flexibility, dynamic control, tuning gain, range of incidence angle, complexity, and the use of an external voltage. The main parameters related to tuning, which are tuning flexibility and the range of the incidence angle, suggest that single convex, parabolic, and spherical lenses are not flexible, hence, steer light from a reduced FoV. However, this resilience is high for RIS elements when compared to a lens with adjustable elements. This allows us to draw the conclusion that single convex, parabolic and spherical lenses, and lens with adjustable elements provide a limited incidence angle range, less than $ 90^o $, while RIS, in addition to their smoothness and accuracy, can steer signals, even at $90^o $ incidence angle. A lens with adjustable elements can be controlled electronically and/or manually, while the RIS structures, presented in this article, are fully tuned electronically. A meta-lens with artificial muscles require higher voltage, up to a couple of kilo-volts \cite{she2018adaptive}, while the crystal-liquid based RIS requires only about 2 V to 5 V to provide a smooth and more efficient variation of the refractive index \cite{sun2019efficient}.
\section{Future Research Opportunities}\label{future}
The application of RIS to VLC systems is novel and requires more investigations. Considering the gains it can provide, it is important that researchers investigate all facets related to the use of RIS in VLC systems. In this sense, it will be important to look at the use of reflect-arrays to steer the VLC beam in the room and analyze the intensity gain obtained while tuning a RIS element using nano-disks with crystal-liquid infiltration, and any other material that is characterized by a low optical opacity, while being electronically tunable.

In VLC, meta-lenses can be exploited to improve signal reception, noise and interference reduction, color selection in multi-wavelength transmissions, and light-beam steering, to enhance the system reliability and quality-of-service. The use of meta-surfaces in VLC may also enable numerous applications in data transfer, road-to-vehicle, device-to-device communications, security, and health-care. These aspects require further studies.

These new technologies can also be exploited in photography where the zoom of the camera will be considerably reduced, allowing the photographers to be closer to their targets and still capture the full image on their screen. The technology can also be combined to a convex lens with adjustable elements to provide short range and higher resolution images. This also requires further studies.    
\section{Conclusion}
This article has elaborated on the use of RIS in VLC systems, to replace the traditional lens and steer the incident light beam. The undeniable advantages of this new technology have been highlighted. RIS helps to improve the VLC receiver's capabilities by increasing its FoV such that it can detect light rays with an incidence angle from 0 to $90^o$, and thus widens the receiver's detection capacity while in rotation. In addition, this article has demonstrated that a substantial intensity gain is obtained at the PD when using RIS elements. Design examples of RIS elements for VLC have also been proposed, and it has been suggested to use meta-lenses with electrically tunable artificial muscles or crystal-liquid cells with embedded nano-disks. Finally, research opportunities towards the use of RIS in VLC systems have been outlined.       
\ifCLASSOPTIONcaptionsoff
\fi
\bibliographystyle{IEEEtran}
\bibliography{A_Manuscript0}
\end{document}